\documentclass[eqsecnum,showpacs,showkeys,floats,aps,nofootinbib,preprint]{revtex4}
\usepackage[latin9]{inputenc}
\usepackage{float}
\usepackage{amsmath}
\usepackage{amssymb}
\usepackage{graphicx}
\usepackage{esint}

\makeatletter
\@ifundefined{textcolor}{}
{%
 \definecolor{BLACK}{gray}{0}
 \definecolor{WHITE}{gray}{1}
 \definecolor{RED}{rgb}{1,0,0}
 \definecolor{GREEN}{rgb}{0,1,0}
 \definecolor{BLUE}{rgb}{0,0,1}
 \definecolor{CYAN}{cmyk}{1,0,0,0}
 \definecolor{MAGENTA}{cmyk}{0,1,0,0}
 \definecolor{YELLOW}{cmyk}{0,0,1,0}
}

\usepackage{bm}\usepackage{amsfonts}\usepackage{mathtools}\usepackage{color}\usepackage{wasysym}\usepackage{latexsym}\usepackage{epsfig}

\makeatother

\begin{document}
\makeatletter

\title{Laser Heater and seeded Free Electron Lasers}

\author{G. Dattoli}

\email{giuseppe.dattoli@enea.it}

\affiliation{ENEA - Centro Ricerche Frascati, via E. Fermi, 45, IT 00044 Frascati
(Roma), Italy}

\author{V. Petrillo}

\email{vittoria.petrillo@mi.infn.it}

\affiliation{INFN - Milano and Università di Milano, via Celoria, 16 20133 Milano,
Italy}

\author{E. Sabia}

\email{elio.sabia@enea.it}

\affiliation{ENEA - Centro Ricerche Frascati, via E. Fermi, 45, IT 00044 Frascati
(Roma), Italy}
\begin{abstract}
In this paper we consider the effect of laser heater on a seeded Free
Electron Laser. We develop a model embedding the effect of the energy
modulation induced by the heater with those due to the seeding. The
present analysis is compatible with recent experimental results and
earlier predictions displaying secondary maxima with increasing heater
intensity. The treatment developed in the paper confirms and extends
the previous analyses and put in evidence further effects which can
be tested in future experiments. 
\end{abstract}
\maketitle

\section{Introduction}

\label{s:intro} Laser Heater (LH) has been proven to be a key element
for Self Amplified Spontaneous Emission Free Electron Laser (SASE-FEL)
operation with high brightness e-beams \cite{Huang}.

The concepts underlying LH trace back to the seminal papers in refs.
\cite{Saldin,Huang2}, where it was pointed out that the onset of
instabilities, like those due to coherent synchrotron radiation (CSR)
\cite{Heifets}, can be counteracted by a FEL type interaction with
an external laser. The induced energy spread reduces the gain of the
instability, thus preventing its growth. The associated physical mechanism
can therefore be ascribed to a manifestation of the Landau damping
, already invoked in the past to account for the experimentally observed
competition between the FEL and the microwave instability in Storage
Ring (SR) devices \cite{Bartolini}. In SR FEL the effect induced
by the laser growth on the microwave instability (and vice-versa)
is fairly complex and can be modeled by merging the equations ruling
the evolution of the two instabilities (FEL and microwave). The final
result is the derivation of a system of non-linear set of differential
equations resembling those of the Volterra prey-predator model, thus
getting a fairly transparent understanding of the dynamics of the
competition. In the case of SR-FEL the mechanism of the competition
is self-regulatory, laser and instability reach a kind of compromise
allowing the coexistence.

In the case of LH, being the laser inducing the beam heating an external
device, the question arises on what should be the amount of laser
intensity to avoid a significant disruption of the quality of the
e-beam, which in turn may prevent the SASE FEL operation.

We expect therefore that, as shown in Fig. 1, the output laser intensity
increases, with the heater power, until the induced Landau damping
is able to control the instability growth beam energy, on the other
side the FEL power decreases when the FEL induced energy spread dominates
the heater process \cite{Dattoli_Labat}.

\begin{figure}[H]
$\qquad\qquad\qquad\qquad\qquad\qquad$\includegraphics[scale=0.4]{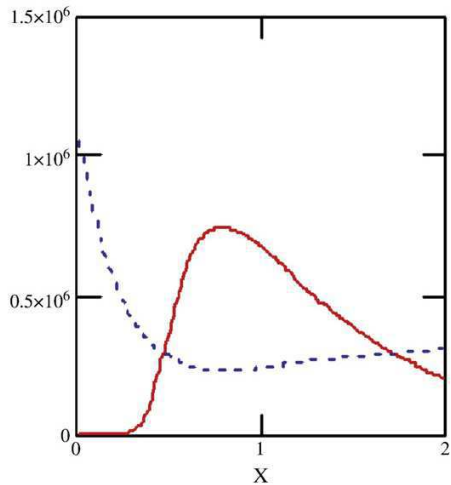}\caption{ FEL intensity (continuous line) and e-beam energy spread (dotted
line) vs. the laser heater power (for the specification of the units
see section III). }
\end{figure}

Recent experimental results \cite{Ferrari} have confirmed such a
``paradigm\textquotedblright{} and have displayed further and interesting
physics when heater is combined with a seeded operation.

The new features emerging from these studies are displayed in Figs.
3 and 4 of ref. \cite{Ferrari} , which show that secondary maxima
appear, for large values of the heater energy.

This effect was predicted in ref. \cite{Saldin}, where it was suggested
that the underlying physical mechanisms are essentially due to the
peculiar nature of the energy distribution acquired by the e-beam
after the heater interaction which does not remain Gaussian anymore.

In this paper we will reconsider the problem ab-initio, by developing
a numerical code based on the solution of the Liouville equation,
which governs the phase space evolution of the e-beam undergoing three
different manipulation stages summarized in Fig. 2: Heating, Acceleration,
Seeding. The novelty of our results relies on the numerical technique
we employ. It allows comprehensive treatment of the problem, yielding
the full dependence of the effect at low and high heater energy and
the inclusion of odd and even harmonics as well. The latter seem characterized
by different behaviors, which could be tested in future experimental
investigations.

The main achievement of the present analysis is the evaluation of
the bunching coefficients, determined in the modulator by the seeding
process. The model we develop has the advantage of reproducing the
behavior at low and large heater energy, by covering correctly the
region before the first peak characterizing the bunching coefficients.

\begin{figure}[H]
$\qquad\qquad\qquad$\includegraphics[bb=0bp 130bp 960bp 550bp,scale=0.45]{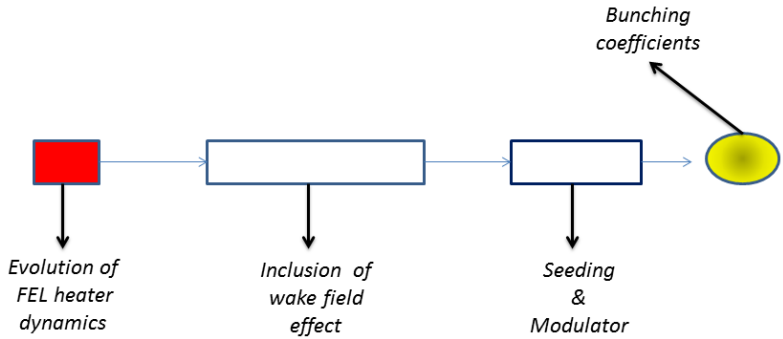}\caption{Flow chart of the procedure adopted }
\end{figure}

\section{The model}

We model the evolution of the FEL-Heater-Seeded device by first solving
the Liouville equation yielding the e-beam phase space distribution
$\rho$ after that the electrons have undergone the heating process,
according to the sketch in Fig. 2.

The Liouville equation we will consider is given below \cite{Dattoli0}

\begin{equation}
\begin{array}{c}
\frac{\partial}{\partial\tau}\rho=-\nu\frac{\partial}{\partial\zeta}\rho+|a|\cos\left(\zeta\right)\frac{\partial}{\partial\nu}\rho,\\
\rho\left(\nu,\zeta\right)|_{\tau=0}=f\left(\nu\right)
\end{array}\label{eq:Liouville equation}
\end{equation}

where $\nu,\zeta$ are the e-beam phase space variable, $\tau$ the
dimensionless time and $\left|a\right|$ is the Colson

FEL dimensionless amplitude \cite{Colson}. The function $f\left(\nu\right)$,
representing the initial condition of our Cauchy problem, is the energy
distribution of the e-beam at the entrance of the heater.

Eq. \eqref{eq:Liouville equation} has been integrated using a simplectic
leapfrog scheme \cite{Dattoli_Sabia} with $f\left(\nu\right)$ being
a Gaussian. Such a functional form is roughly preserved for modest
values of the heater energy, but, when it increases, the distribution
is distorted and secondary lobes appear \cite{Colson} (see Fig. 3).

Being not interested into the details of the CSR inside the LINAC,
we use a standard procedure (described e. g. in \cite{Saldin}), modelling
the effect of the heater through the wake field induced energy spread.

The total energy spread (CSR + Heater) vs. the heater energy, at the
entrance of the modulator, is shown in Fig. 4, the behaviour is easily
understandable. After a damping of the instability, the laser induced
energy spread becomes dominant and the total energy spread increases.

\begin{figure}[H]
$\qquad\qquad\qquad\qquad\qquad$\includegraphics[scale=0.35]{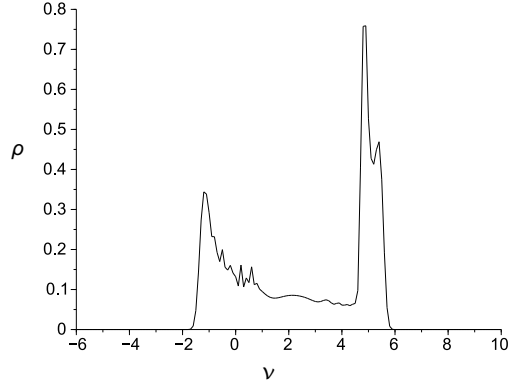}

\caption{Energy distribution distortion at $\left|a\right|^{2}=16$}
\end{figure}

\begin{figure}[H]
$\qquad\qquad\qquad\qquad\qquad$\includegraphics[scale=0.35]{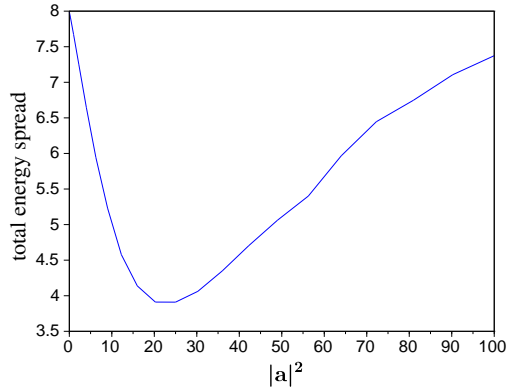}\caption{Total energy spread (CSR+ Heater) vs. $\left|a\right|^{2}$.}
\end{figure}

We have modelled the e-beam phase space evolution in the modulator
using the same Liouville equation in eq. \eqref{eq:Liouville equation},
in which $\left|a\right|$ is replaced by the seed amplitude and the
phase space variables are rescaled according to the new parameters
(the average beam energy and the undulator geometry). The energy distribution
is assumed to be that calculated with the combined effect of Heater
and CSR. The previously adopted solution procedure of the Liouville
equation yields the e-beam phase space distribution at the entrance
of the modulator. We can therefore expand the phase space distribution
in a Fourier series, namely

\begin{equation}
\rho\left(\nu,\zeta\right)=\sum_{n}b_{n}e^{in\zeta}\label{eq:Fourier series}
\end{equation}

and then we evaluate the bunching coefficients $b_{n}$ as

\begin{equation}
b_{n}\left(\nu\right)=\frac{1}{2\pi}\int_{0}^{2\pi}\rho\left(\nu,\zeta\right)e^{-in\zeta}d\zeta\label{eq:bunching coefficients}
\end{equation}

By taking a further average on the energy distribution, we obtain
what is shown in Figs. 5, regarding the energy averaged bunching coefficients
($n=3,\,13$) vs. the heater energy. The case n=2 is reported in Fig.
6. The horizontal axis in the previous figures is expressed in terms
of the dimensionless amplitude $\left|a\right|^{2}$ , which is in
turn related to the heater energy. We have used dimensionless unit
rather than dimensional because we prefer to provide general trends
which can be later adapted to a more specific experimental cofiguration
as we will see in the following. A rescaling of the dimensionless
amplitude in terms of heater energy will be discussed in the concluding
section of this paper. 

The behavior is paradigmatically similar for all the harmonics: a
growth, a maximum, a decrease, new secondary maxima. The larger harmonic
number corresponds to a more significant sensitivity of the maxima
to the heater energy, which is reflected by a narrower peak and by
a smaller peak to peak ratio.

The even harmonic $n=2$ displays a more intrigued pattern providing
a secondary maximum larger than the first.

The first maxima are obtained in correspondence of the minimum of
the total energy spread (Fig. 6), the secondary maxima are associated
with a more complicated dynamical behaviour we will discuss below.

\begin{figure}[H]
\includegraphics[scale=0.33]{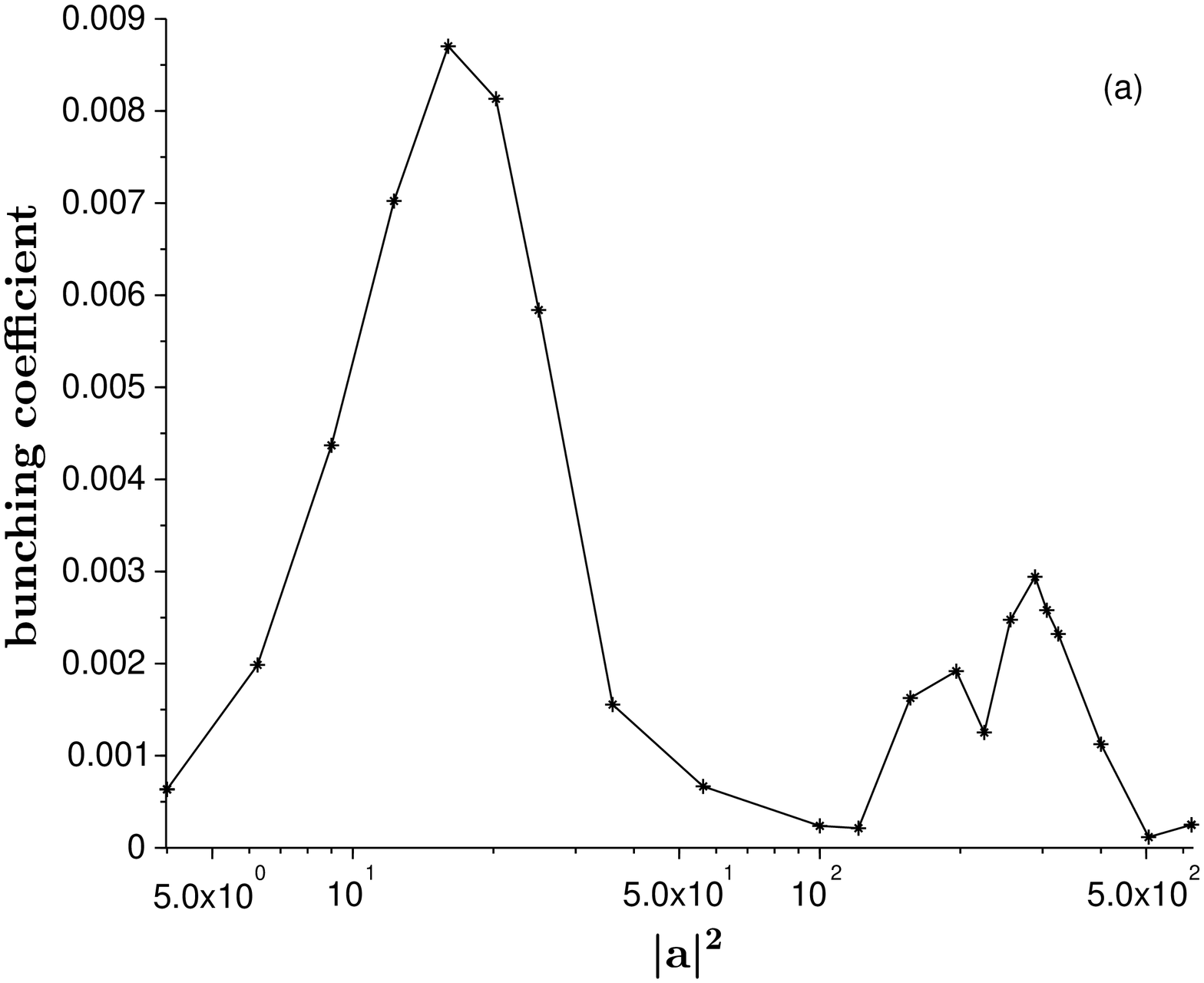}$\:$\includegraphics[scale=0.33]{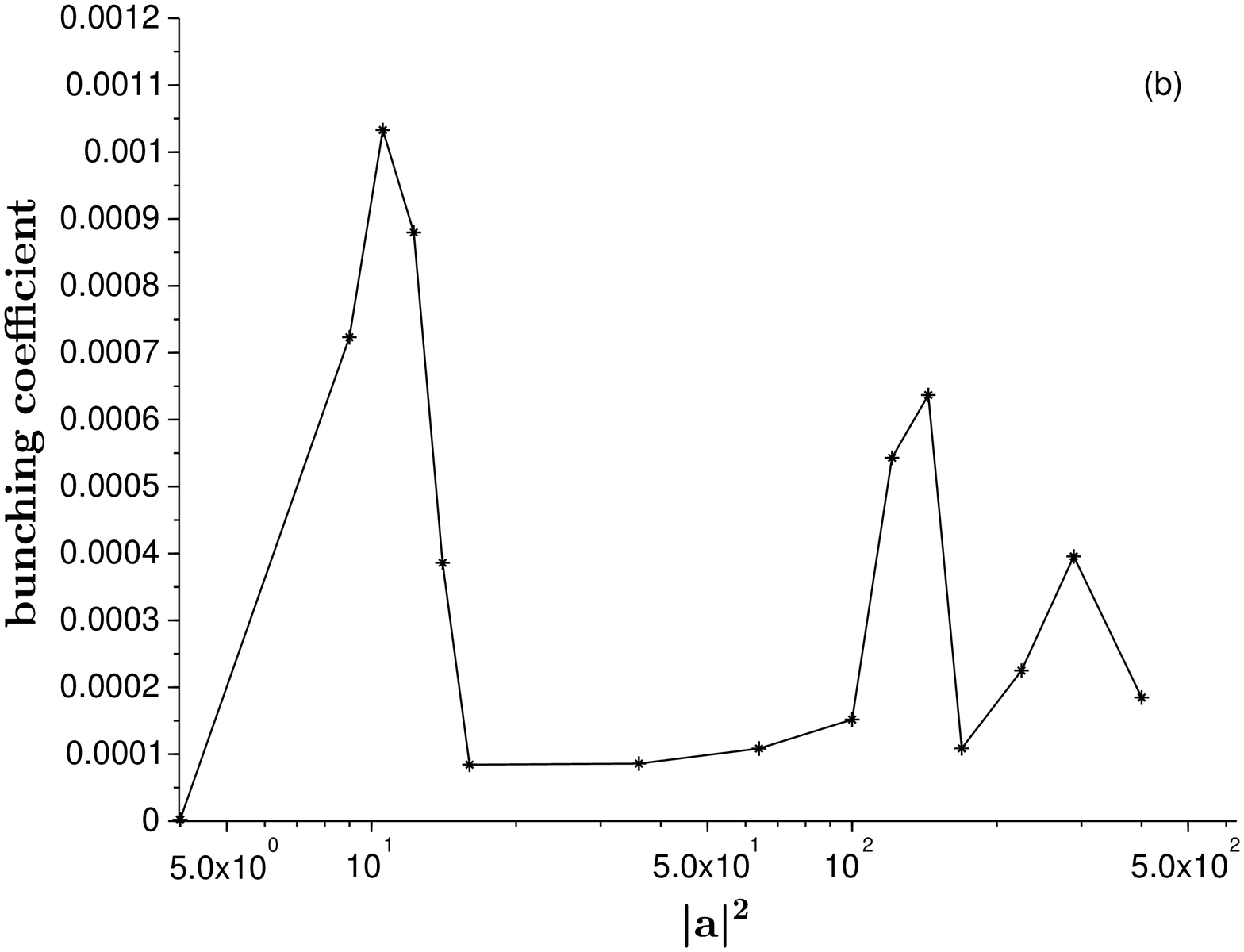}\caption{Bunching coefficients at the end of the modulator vs. $\left|a\right|^{2}$,
(a) n=3, (b) n=13.}
\end{figure}

\begin{figure}[H]
$\qquad\qquad\qquad\qquad\qquad$\includegraphics[scale=0.35]{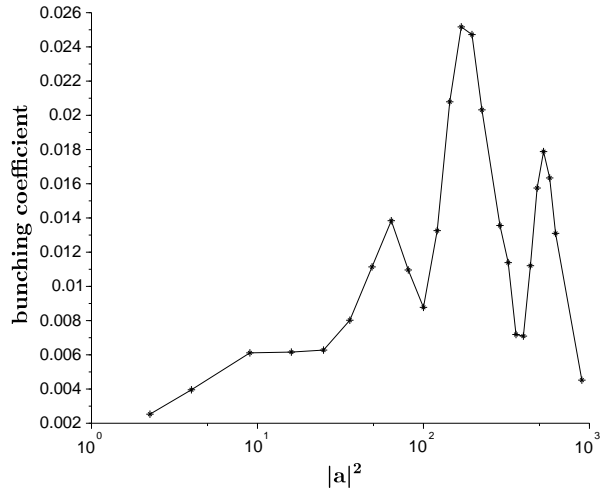}\caption{Same of Fig. 5 for n=2}
\end{figure}

In Fig. 7 we show the position of the maxima and the total energy
spread vs. the heater energy. It is worth stressing that the first
maxima of the odd harmonics are all in correspondence (even though
not exactly the same) of the minimum of the total energy spread, the
secondary maxima (including those of the second harmonic) are shifted
towards the region in which the heater induced energy spread is dominating
with respect to the CSR spread. It must be stressed that in this region
of the heater energy, the energy distribution at the entrance of the
modulator is far from being a Gaussian and the energy spread (considered
as the r.m.s. of the distribution) cannot be considered fully representative
of the whole distribution.

\begin{figure}[H]
$\qquad\qquad\qquad\qquad\qquad$\includegraphics[scale=0.4]{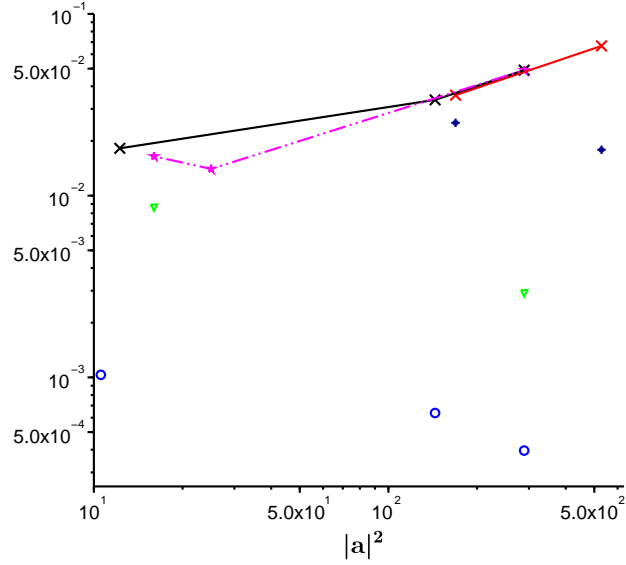}\caption{Bunching coefficient peak position n=3 (green triangle mark), n=13
(blue circle mark), n=2 (black diamond mark) vs. $\left|a\right|^{2}$,
the interpolating line refers to the heater induced energy spread }
\end{figure}

In Figs. 9-10 we have reported the energy distribution at the entrance
of the modulator and at the end of the modulator for different values
of the dimensionless seeding amplitude $\left|a\right|$, corresponding
to the position of the first and second maxima of the bunching coefficient
$n=3$ .

\begin{figure}[H]
\includegraphics[scale=0.35]{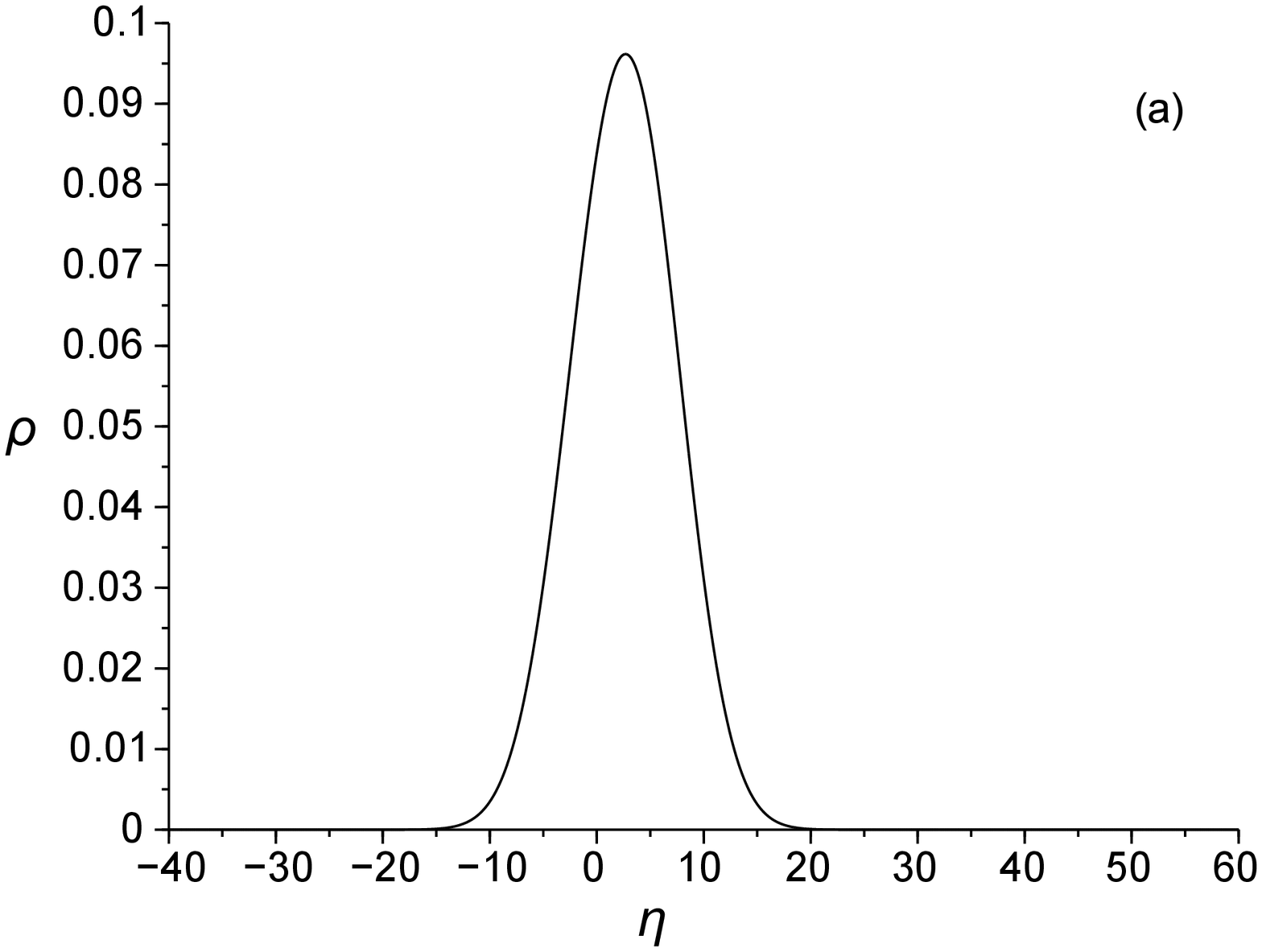}\includegraphics[scale=0.35]{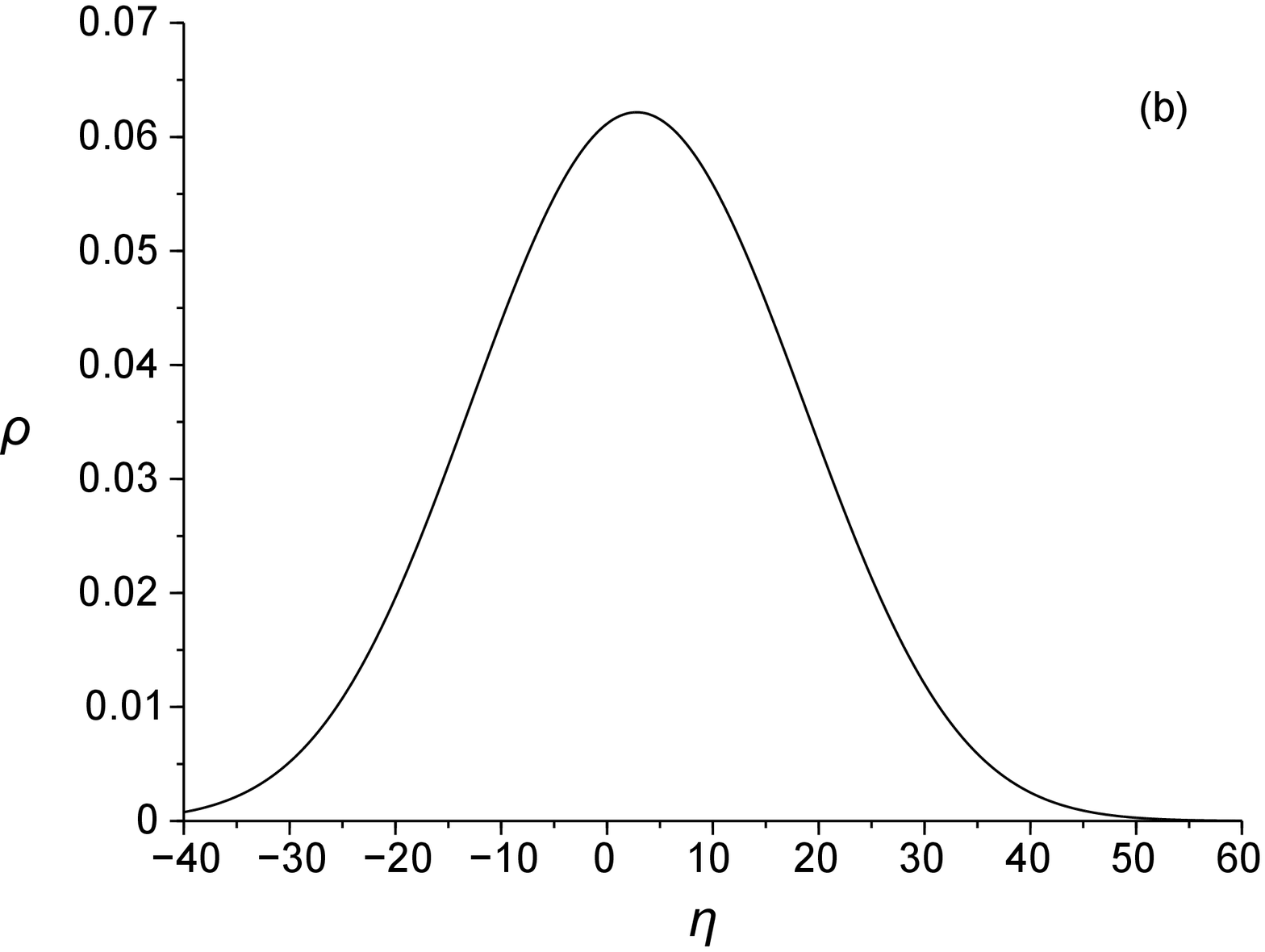}\caption{Energy distribution at the entrance of the modulator for (a) $\left|a\right|^{2}=16$,
(b) $\left|a\right|^{2}=289$. }
\end{figure}

\begin{figure}[H]
\includegraphics[scale=0.35]{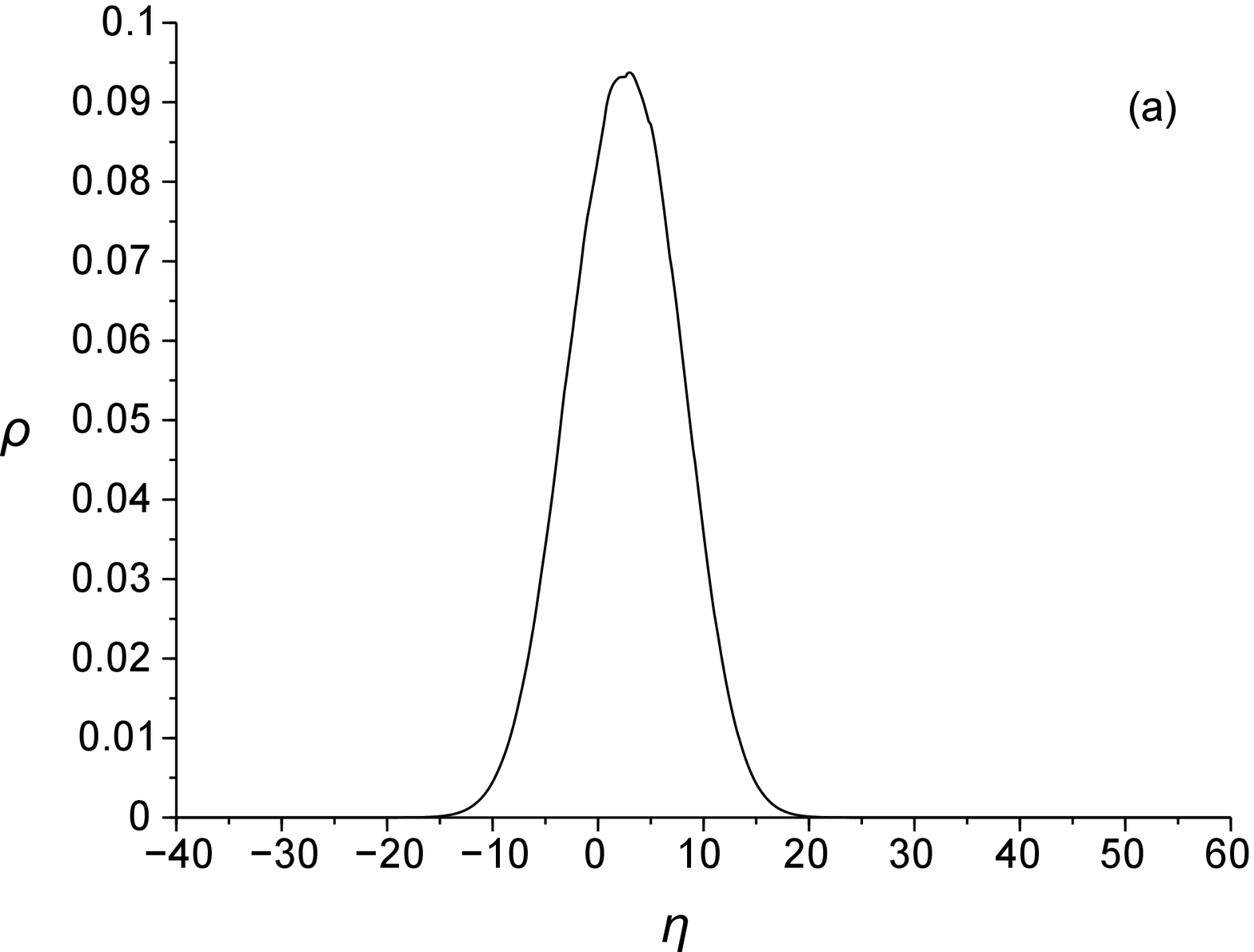}\includegraphics[scale=0.35]{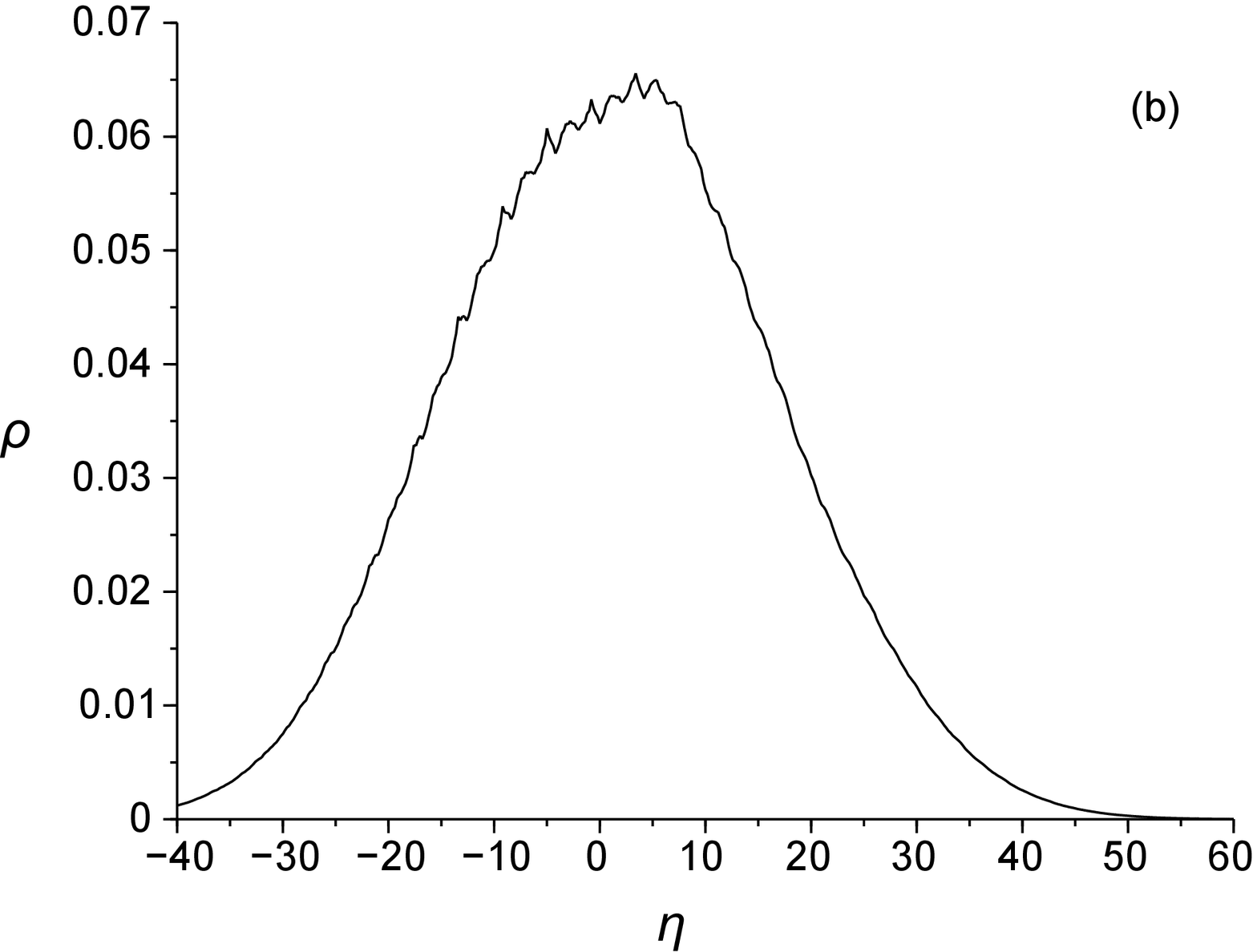}\caption{Energy distribution at the end of the modulator for (a) $\left|a\right|^{2}=16$,
(b) $\left|a\right|^{2}=289$. }
\end{figure}

\section{Final Comments}

The results obtained so far confirm that the consequences of the heater
interaction are more rich than usually believed \cite{Huang2,Ferrari}.
It seems indeed that the successive passage of the beam inside the
Linac accelerating sections is such that it preserves some memory
of the effects by the FEL dynamics inside the first undulator. The
heater does not simply induces a Gaussian noise, but has non-trivial
effects which mark the successive emission inside the modulator in
a way reminiscent of the echo mechanism \cite{Stupakov}.

The Colson dimensionless amplitude $\left|a\right|$ is associated
with the laser intensity $I$ according to the identity \cite{Dattoli0,Ciocci}%
\footnote{Even though the laser used in the heater is an external laser and
not a FEL, we use the same formalism of self-induced heating because
the effects are the same.%
}

\begin{equation}
\begin{array}{c}
\left|a\right|^{2}=0.8\pi^{4}X\\
X=\frac{I}{I_{s}}
\end{array}\label{eq:relation}
\end{equation}

where $I_{s}$ is the FEL saturation intensity, which is in turn linked
to the beam and undulator parameters by

\begin{equation}
\begin{array}{c}
I_{s}\left[\frac{MW}{cm^{2}}\right]\cong6.9\times10^{2}\left(\frac{\gamma}{N}\right)^{4}\frac{1}{\left[\lambda_{u}\left[cm\right]Kf_{b}\left(\xi\right)\right]^{2}},\\
f_{b}\left(\xi\right)=J_{0}\left(\xi\right)-J_{1}\left(\xi\right),\quad\xi=\frac{1}{4}\frac{K^{2}}{1+\frac{K^{2}}{2}}
\end{array}\label{eq:FEL saturation intensity}
\end{equation}

Using the parameters reported in ref. \cite{Ferrari} for the heater
section ($\gamma\cong200,\;\lambda_{u}=4cm,\; K=1.17,\: N=12$) we
obtain for the saturation intensity of the heater section $I_{s}\cong3\times10^{6}\frac{MW}{cm^{2}}$.
Using such a value as reference number we obtain that in, correspondence
of the first maximum of the third harmonic bunching coefficient, the
corresponding laser intensity is $I\cong\frac{\left|a\right|^{2}}{0.8\pi^{4}}I_{s}=0.6\times10^{6}\frac{MW}{cm^{2}}$.

A comparison between the results presented in Figs. 5 and those reported
in ref. \cite{Ferrari} in terms of absolute energy values is made
difficult, since our treatment does not include 3-D effects and we
cannot determine the effective overlapping between electron and photon
beams. A comparison is however possible if we adopt a different strategy.
We ``calibrate'' the horizontal axis in Fig. 5 in such a way that
the first maximum coincides with $1\mu J$. By overlapping the plot
in Fig. 5a, with the experimental results concening the FEL intensity
at 32 nm in the radiator for the case of FERMI experiment, we obtain
what has been reported in Fig. 10, the comparison is sasfactory, even
though our calculation refers to the square modulus of the bunching
coefficient at the end of the modulator and the experimental results
to the FEL intensity at the end of three undulator sections in the
radiator. in these conditions, being eliminated the gain effects,
the output pulse intensity is directly proportional to the square
modulus of the bunching coefficient itself. Therefore we normalize
the peak bunching with the peak energy reported in ref. {[}7{]}. An
important remark is that the analysis developed in this paper correctly
predicts the low LH intensity behaviour and not only that at larger
energy. To make the comparison more subsantive we have reported in
Fig. 11 the slice induced energy spread, in the heater section: we
have interpolated our numerical results with the analytical formula
\cite{Dattoli_Ottaviani}

\begin{equation}
\sigma(X)=\frac{0.433}{N}\: e^{-0.25\beta X+0.01\beta^{2}X^{2}}\sqrt{\frac{\beta X}{1-e^{-\beta X}}-1}\label{eq:slice induced energy spread}
\end{equation}

The comparison with the experimental results quoted in \cite{Spampinati}
provides quite a good agreement, which has been further checked using
the numerical procedure based on the FPE equation.

\begin{figure}[H]
$\qquad\qquad\qquad\qquad\qquad$\includegraphics[scale=0.35]{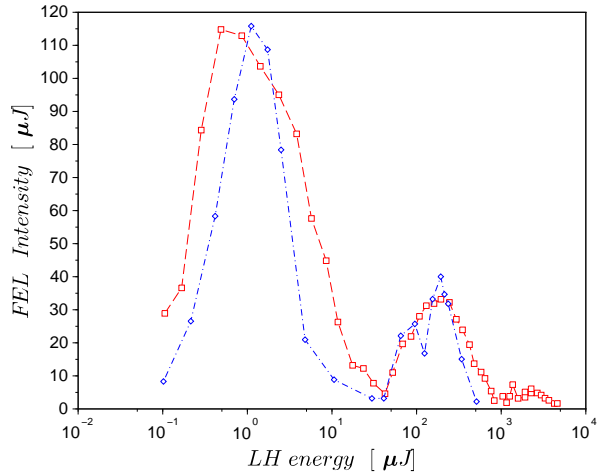}

\caption{Comparison between the results presented in Fig. 5a (blue diamond
dashed-dot) and those reported in ref. \cite{Ferrari} (red box dashed). }

\end{figure}

\begin{figure}[H]
$\qquad\qquad\qquad\qquad$\includegraphics[scale=0.35]{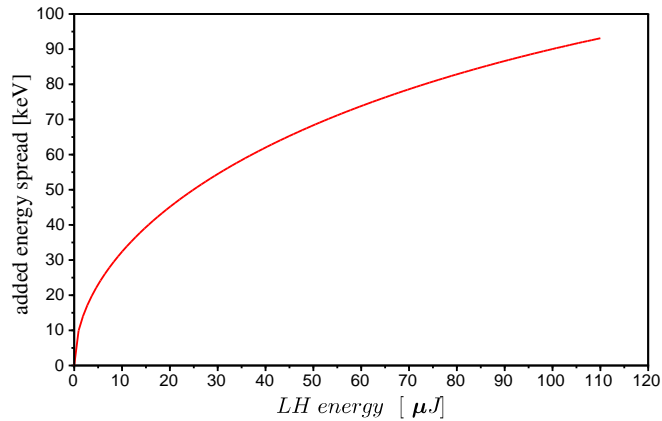}\caption{The slice induced energy spread, in the heater section vs. LH energy}

\end{figure}

Since the amount of laser energy quoted in the experiment for the
first peak is around $E_{L}\cong1\mu J$ , according to the characteristic
of the laser for the FERMI heater reported in \cite{Ferrari} (namely
laser pulse duration $\tau_{L}\cong10ps$, transverse section $\sigma_{L}\cong0.2mm$
) we can evaluate the laser pulse intensity corresponding to $I_{L}=\frac{E_{L}}{2\pi\sigma_{L}^{2}\tau}\cong0.6\frac{MW}{cm^{2}}$.
The second peak is predicted by our model to be determined by a laser
energy larger by a factor 20 than that corresponding to the first.

The agreement between theory and experiment can be considered satisfactory.

We must underline that our treatment holds for large laser size compared
to the e-beam, namely for $\sum_{L}\gg\sum_{E}$ , when such a condition
does not occur the energy distribution after the heater is more similar
to a Gaussian. This aspect of the problem has however been thoroughly
discussed in ref. \cite{Huang2} and will not reconsidered here.

The analysis we have developed in the paper is straightforward from
the conceptual point of view but rather heavy in terms of computer
time.

To reduce the number of calculations we have simplified the part concerning
the effect of the Landau Damping inside the Linac. We have assumed
indeed a heuristic model implying that the heater energy induced distribution
combines with that due to the wake field inside the Linac structures,
becoming less efficient in providing additional spread with increasing
heating noise. The model is essential that proposed in ref. \cite{Saldin}.

It has been stressed that the occurrence of new maxima emerging in
the bunching coefficients is associated with the fact that the energy
distribution deviates from a Gaussian after the heater. We have considered
the dependence on the heater energy of the average energy value, rms
and of the Fisher index \cite{Fisher} defined as $\delta=\frac{m_{4}-3\sigma^{4}}{\sigma^{4}}$
(see Fig. 12). This last quantity measures the relevance of the dissymmetry
and indeed it becomes more negative with increasing heater energy,
as also evident by comparison with Fig. 9.

\begin{figure}[H]
$\qquad\qquad\qquad\qquad\qquad$\includegraphics[scale=0.35]{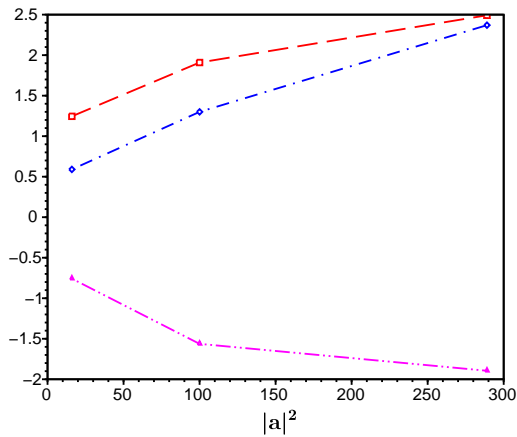}\caption{First moment of the energy distribution (red square mark), second
moment (blue diamond mark), Fisher parameter (magenta triangle mark).
The second moment (namely the induced energy spread) has been divided
by 10.}
\end{figure}

The analysis we have developed agrees fairly well with the experimental
results even though the modelling of the heater is rather simplified.
We have used fairly general method which predicts that the effect
of the recurrence of maxima is present also for higher order bunching
coefficients, even though the peaks are significantly narrower. We
also underline that the present analysis satisfactorily reproduces
the whole trend of the curves from low to high heater energy. It is
also worth stressing that the position of the peaks does not change
dramatically with the heater energy (at least for odd harmonics).
However, regarding the case of even order bunching coefficients, it
seems that the structure of the peaks is different from that of the
odd cases with the secondary maxima more pronounced and shifted towards
higher energy values. At the moment we have no convincing explanation
for such a behavior, which, if confirmed by the experiment, is still
an open question.

\textbf{\large{{{Acknowledgements}}}}{\large \par}

The Authors express their sincere appreciation to L. Giannessi for
a clarifying discussion on the results of ref. \cite{Ferrari} and
to T. Dubuc for the analysis of the experimental data.


\end{document}